
\input harvmac
\noblackbox
\def\Title#1#2{\rightline{#1}\ifx\answ\bigans\nopagenumbers\pageno0\vskip1in
\else\pageno1\vskip.8in\fi \centerline{\titlefont #2}\vskip .5in}

scaled\magstep3
 
scaled\magstep3
%

\def\ajou#1&#2(#3){\ \sl#1\bf#2\rm(19#3)}
\def\frac#1#2{{#1 \over #2}}

\def\mn{{\mu\nu}}
\def\a{\alpha}

\def\t{\theta}
\def\p{\partial}
\def\to{{\rightarrow}}
\def\vp{{\varphi}}
\def\({\left (}
\def\){\right )}
\def\[{\left [}
\def\]{\right ]}
\def\KK{{Kaluza-Klein}}
\def\mam{{monopole-antimonopole}}
\def\r{r_H}
\def\q{{\bf q}}
\def\S{\Sigma}

\def\tvp{{\widetilde{\varphi}}}
\def\tr{{\widetilde r}}

\def\R{{\bf R}}

%
%
\def\npb#1#2#3{{\sl Nucl. Phys.} {\bf B#1} (#2) #3}

%
%

\lref\gl{R. Gregory and R. Laflamme,
\ajou Phys. Rev. Lett. &70 (93) 2837;\ajou Nucl. Phys. &B428 (94) 399.}
\lref\polchinski{J. Polchinski, \ajou Phys. Rev. Lett &75 (95) 4724.} 
\lref\bgv{R. Basu, A. Guth and A. Vilenkin, \ajou Phys. Rev. &D44 (91) 340.}
\lref\all{
{G.W. Gibbons,
in {\sl Fields and geometry}, proceedings of
22nd Karpacz Winter School of Theoretical Physics: Fields and
Geometry, Karpacz, Poland, Feb 17 - Mar 1, 1986, ed. A. Jadczyk (World
Scientific, 1986);\hfil\break
D. Garfinkle and A. Strominger,
\ajou Phys. Lett. &256B (91) 146;\hfil\break
D. Garfinkle, S. Giddings and A. Strominger, \ajou Phys. Rev. &D49 (94) 958;
\hfil\break
S.F. Ross, \ajou Phys. Rev. &D49 (94) 6599; \hfil\break
S.W. Hawking, G.T. Horowitz and S.F. Ross, \ajou
Phys. Rev. &D51 (95) 4302;\hfil\break
P. Yi, \ajou Phys. Rev. &D51 (95) 2813; ``Quantum Stability of
Accelerated Black Holes," hep-th/9505021\semi
D. Eardley, G. Horowitz, D. Kastor, J. Traschen,
Phys. Rev. Lett. {\bf 75} (1995) 3390,
gr-qc/9506041\semi
S.W. Hawking and S.F. Ross,  \ajou Phys. Rev. Lett. &75 (95) 3382;
\hfil\break
R. Emparan, \ajou Phys. Rev. Lett. &75 (95) 3386;\hfil\break
R. Gregory and M. Hindmarsh, Phys. Rev. {\bf D52} (1995) 5598\semi
S. W. Hawking, ``Virtual Black Holes," hep-th/9510029.
}}
\lref\sen{A. Sen, \ajou Mod. Phys. Lett. &A10 (95) 2081.}
\lref\bbs{K. Becker, M. Becker and A. Strominger, ``Fivebranes, membranes
and nonperturbative string theory," hep-th/9507158.}
\lref\AfMa{I.K. Affleck and N.S. Manton,
\ajou Nucl. Phys. &B194 (82) 38\semi
I.K. Affleck, O. Alvarez, and N.S. Manton,
\ajou Nucl. Phys. &B197 (82) 509.}
\lref\preskill{S. Coleman, J. Preskill, and F. Wilczek, \ajou Nucl. Phys.
&B378 (92) 175.}
\lref\brown{J.D. Brown, ``Pair Creation of Electrically Charged Black
Holes,'' gr-qc/9412018.}
\lref\brownt{ J. Brown and C. Teitelboim, \ajou  Nucl. Phys. &B297 (88) 787;
J. Garriga, \ajou Phys. Rev. &D49 (94) 6327.}
\lref\hrdual{S.W. Hawking and S.F. Ross, \ajou Phys. Rev. &52 (95) 5865.}
\lref\ross{{S.F. Ross, \ajou Phys. Rev. &D49 (94) 6599.}}
\lref\emparan{R. Emparan, ``Pair Creation of Black Holes Joined
by Cosmic Strings,'' gr-qc/9506025.}
\lref\rosstwo{S.F. Ross,
``Pair Creation Rate for $U(1)^2$ Black Holes,''
gr-qc/950910.}
\lref\hhr{S.W. Hawking, G.T. Horowitz and S.F. Ross, \ajou
Phys. Rev. &D51 (95) 4302.}
\lref\gh{R. Gregory and M.B. Hindmarsh ``Smooth Metrics for Snapping
Strings,'' gr-qc/9506054.}
\lref\ehkt{D. Eardley, G. Horowitz, D. Kastor, J. Traschen, ``Breaking
Cosmic Strings without Monopoles.'' gr-qc/9506041.}
\lref\decay{ H.F. Dowker, J.P. Gauntlett,  G.W. Gibbons, and G.T. Horowitz,
``The decay of magnetic fields in Kaluza-Klein theory," hep-th/9507143.}
\lref\pagepope{
D.N. Page and C.N. Pope, \ajou Class. Quant Grav & 3 (86) 249-259}
\lref\gibhaw{G. Gibbons and S. Hawking, \ajou Commun. Math. Phys.
&66 (79) 291-310.}
\lref\baal{ P. van Baal, F.A. Bais and P van Niewenhuizen, ajou Nucl. Phys. &B
233 (84) 477. }
\lref\ramond{P. Ramond, in {\it Lattice Gauge Theories, Supersymmetry,
and Grand Unification: Proceedings of 6th Workshop on Current Problems in
High-Energy Particle Theory}, Johns Hopkins University (1982).}
\lref\chgr{J. Cheeger and D. Gromoll, \ajou J. Diff. Geom. &6 (71) 119-128.}
\lref\bergery{ L Berard Bergery {\sl Inst Elie Cartan } {\bf 6} (1982) 1-60 }
\lref\baisbat{F. Bais and P. Batenberg, {\sl Nucl. Phys.} {\bf B 253}
(1985) 162-172.}
\lref\pptwo{D N Page \& C N Pope {\sl Class Quant Grav} {\bf 4}
(1987) 213-225.}
 \lref\xuli{J Xu \& X Li {\sl Phys Lett } {\bf B 208 } (1988) 391-395}
\lref\ezawakoh{ Z F Ezawa \& I G Koh {\sl Phys lett} {\bf B 140} (1984)
205-208}
\lref\kaolee{H-C. Kao \& K. Lee, \ajou  Phys. Rev. &D52 (95) 6050.}
\lref\liyuzhang{ X Li , F Yu, \& J-Z Zhang {\sl Phys Rev }
{\bf D 34} (1986) 1124-1128}
\lref\barr{  S M Barr, {\sl Phys Lett } {\bf B 129} (1983) 303-306}
\lref\perry{ M J Perry, {\sl Phys Lett } {\bf B 137} (1983) 171-174}
\lref\yuille{ A Yuille, {\sl Class Quant Grav} {\bf 4} (1987) 1409-1426}
\lref\chakra{ A Chakrabarti, {\sl Phys Lett } {\bf B 172} (1986) 175.}
\lref\young{ R E Young, {\sl Phys rev } {\bf D 28} (1983) 2420-2430}
\lref\dghr{ A. Dabholkar, G. Gibbons, J. A. Harvey and F. R. Ruiz-Ruiz,
 \npb{340}{1990}{33}.}
\lref\dggh{H.F. Dowker, J.P. Gauntlett, S.B. Giddings and G.T. Horowitz,
\ajou Phys. Rev. &D50 (94) 2662.}
\lref\dgkt{H.F. Dowker, J.P. Gauntlett, D.A. Kastor, J. Traschen,
\ajou Phys. Rev. &D49 (94) 2909.}
\lref\andy{D. Garfinkle, S. Giddings and A. Strominger, ``Entropy in Black
Hole Pair Production,'' Santa Barbara preprint UCSBTH-93-17,
gr-qc/9306023.}
\lref\gm{G.W. Gibbons and K. Maeda,
\ajou Nucl. Phys. &B298
(88) 741.}
\lref\ghs{D. Garfinkle, G. Horowitz, and A. Strominger,
\ajou Phys. Rev. &D43 (91) 3140, erratum\ajou Phys. Rev.
& D45 (92) 3888.}
\lref\gwg{G.W. Gibbons,
in {\sl Fields and geometry}, proceedings of
22nd Karpacz Winter School of Theoretical Physics: Fields and
Geometry, Karpacz, Poland, Feb 17 - Mar 1, 1986, ed. A. Jadczyk (World
Scientific, 1986).}
\lref\garstrom{D. Garfinkle and A. Strominger,
\ajou Phys. Lett. &256B (91) 146.}
\lref\ernst{F. J. Ernst, \ajou J. Math. Phys. &17 (76) 515.}
\lref\rafkk{R. D. Sorkin, \ajou Phys. Rev. Lett. &51 (83) 87.}
\lref\sorkin{R. D. Sorkin, \ajou Phys. Rev. &D33 (86) 978.}
\lref\grossperry{D. Gross and M.J. Perry, \ajou Nucl. Phys. &B226 (83) 29.}
\lref\schwinger{J. Schwinger}
\lref\melvin{M. A. Melvin, \ajou Phys. Lett. &8 (64) 65.}
\lref\mp{R. C. Myers and M. J. Perry,
\ajou Ann. Phys. &172 (86) 304.}
\lref\witten{E. Witten, \ajou Nucl. Phys. &B195 (82) 481.}
\lref\ginsperry{P. Ginsparg and M.J. Perry, \ajou Nucl.Phys. &B222 (83) 245.}
\lref\gott{J.R. Gott, \ajou Nuov. Cim. &22B (74) 49.}
\lref\schulman{L.S. Schulman, \ajou Nuov. Cim. &2B (71) 38.}
\lref\peres{A. Peres, \ajou Phys. Lett &31A (70) 361.}
\lref\townsend{P. Townsend, \ajou Phys. Lett. &B350 (95) 184. }
\lref\hulltown{ C. Hull and P. Townsend, \ajou Nucl. Phys. &B438 (95) 109.}
\lref\wittena{E. Witten, \ajou Nucl. Phys. &B443 (95) 85.}
\lref\ghl{J. P.~Gauntlett, J. A. Harvey, and J. T.~Liu,
\npb{409}{1993}{363}.}
\lref\khuri{R. Khuri, \npb{387}{1992}{315}.}
\lref\strom{A. Strominger, \ajou Nucl. Phys. &B451 (95) 96.}
%
%
\Title{\vbox{\baselineskip12pt
\hbox{CALT-68-2031}
\hbox{DAMTP-R/95/57}
\hbox{UCSBTH-95-37}
\hbox{hep-th/9512154}}}
{\vbox{\centerline{Nucleation of $P$-Branes and Fundamental Strings}}
}
{
\baselineskip=12pt
\centerline{Fay Dowker$^{1a}$, Jerome P. Gauntlett$^{1b}$,
Gary W. Gibbons$^{2}$ and Gary T. Horowitz$^{3}$}
\bigskip
\centerline{\sl $^1$California Institute of Technology}
\centerline{\sl Pasadena, CA, 91125}
\centerline{\it $^a$dowker@theory.caltech.edu,  $^b$jerome@theory.caltech.edu}
\medskip
\centerline{\sl $^2$DAMTP, University of Cambridge}
\centerline{\sl Silver St., Cambridge, CB3 9EW}
\centerline{\it gwg1@anger.amtp.cam.ac.uk.edu}
\medskip
\centerline{\sl $^3$Department of Physics}
\centerline{\sl University of California}
\centerline{\sl Santa Barbara, CA 93106}
\centerline{\it gary@cosmic.physics.ucsb.edu}
\bigskip
\smallskip
\centerline{\bf Abstract}
We construct a solution to the low-energy string equations of
motion in five dimensions that describes a circular loop
of fundamental string exponentially expanding in a background
electric $H$-field. Euclideanising this gives an instanton for
the creation of a loop of fundamental string in a background
$H$-field, and we calculate the rate of nucleation. Solutions
describing magnetically charged strings and $p$-branes, where
the gauge field comes from Kaluza-Klein reduction on a circle,
are also constructed. It is known that a magnetic flux tube in
four (reduced) spacetime dimensions is unstable to the pair creation
of Kaluza-Klein monopoles. We show that in $(4+p)$ dimensions,
magnetic $(p+1)$ ``fluxbranes" are unstable to the nucleation of a
magnetically charged spherical $p$-brane. In ten dimensions the
instanton describes the nucleation of a Ramond-Ramond magnetically
charged six-brane in type IIA string theory. We also find static
solutions describing spherical charged $p$-branes or fundamental
strings held in unstable equilibrium in appropriate background fields.
Instabilities of intersecting magnetic fluxbranes are also discussed.

\Date{12/95}
\vfill\eject
%

\newsec{Introduction}

Solitons have played an important role in several recent developments
in string theory. In particular, they appear to be a key to understanding
various nonperturbative aspects of the theory \refs{\hulltown,\wittena,\strom,
\polchinski}.
Surprising connections have been found between string states
and black holes
and between strings and higher dimensional
extended objects, $p$-branes.
There are indications that these objects all
play a fundamental
role in the theory.

It has been shown that localized solitons such
as monopoles can be pair created in a background
magnetic field \AfMa. Recently there has been considerable
interest in the analogous process involving gravity:
the pair creation of charged black holes in
background electromagnetic fields and by breaking
cosmic strings \refs{\all, \dgkt,\dggh,\decay,\brown,\hrdual}.
The question naturally
arises as to whether extended objects such as $p$-branes
and fundamental strings can
also be produced quantum mechanically in appropriate background fields.
The special case of $p$-branes
coupled to a, cosmological constant inducing, $(p+1)$-form potential
in $(p+2)$ spacetime
dimensions was previously discussed  in \brownt. The nucleation 
of vortex loops, local and global,
has also been investigated in four dimensions (see \refs{\kaolee} and 
references therein).

We will present a solution to the low-energy string equations of
motion that describes a finite
loop of fundamental string in five spacetime dimensions
expanding in a background electric-type
$H_{\mu\nu\rho}$ field.
Analytically continuing this solution
yields an instanton corresponding to the nucleation of a single loop
of fundamental string.
We also find related solutions in $(p+4)$ spacetime
dimensions that describe spherical, magnetically charged
$p$-branes expanding in a background magnetic field.
Again, analytically continuing the expanding
solution gives an instanton for the nucleation of a charged $p$-brane.
Along the way we will construct static versions of the Lorentzian
solutions: a loop of fundamental string or spherical magnetic
$p$-brane held in unstable equilibrium in a background field.
A further generalization results in solutions describing
spherical uncharged branes of any odd (even) dimension
either in static unstable equilibrium
or expanding in background magnetic fields in odd (even) spacetime
dimensions.

The construction of these solutions relies
on three observations.
We begin by considering
Kaluza-Klein theory with a $U(1)$ reduction
from $D$ spacetime dimensions to $D-1$. 
The first observation is that the spatial
part of a basic
\KK\ monopole \refs{\rafkk,\grossperry}
can be locally constructed by taking ${\bf R}^4$
and considering the $U(1)$ isometry that
simultaneously rotates the two orthogonal two-planes by the
same angle. This acts freely except for a fixed point at the origin.
Dividing out
by this action gives us a configuration in three spatial dimensions
that is, locally, the \KK\ monopole. Magnetically charged
higher $p$-branes
are given by multiplying this, locally, by $p$ extra
trivial directions so that the fixed point set of the induced rotation
in $\R^{4+p}$
is the brane. $p$-branes that are not magnetically charged
are formed locally by taking the quotient of ${\bf R}^{2k+p}$ by
rotations that simultaneously rotate in $k\ne 2$
two-planes. The $p$-brane is the fixed point set
of the rotation. Only in
$(p+4)$ spacetime dimensions can the $p$-brane
be charged with respect to the two-form Maxwell field, $F$, arising from
a circle reduction.
The general fixed point set analysis will be described in
detail in section 2.

The second ingredient in our construction is a
better understanding of the pair-production of $D=5$ Kaluza-Klein
monopoles in magnetic flux tubes that
has recently been gained \refs{\decay}.
Remarkably this process has been shown to be closely related to
an instability of \KK\ magnetic fields analogous to that
of the ordinary \KK\ vacuum described by Witten \refs{\witten}.
It has been found that not only the
topology but also the metric of the instantons involved is the same
in both cases. The topology is ${\bf R}^2 \times S^3$
and the metric is that of the five-dimensional
rotating  black hole discovered by Myers and Perry \mp.
The idea that emerged from this work is that the monopoles that
are pair produced via this instanton arise as the fixed points
of the  $U(1)$ isometry that one divides out by in performing the
\KK\ reduction. Thus, to construct the higher dimensional generalizations
given here,
we take higher dimensional black holes and divide out by an
appropriate $U(1)$ action such that the fixed point sets
are the desired $p$-branes.

Since these are solutions to the vacuum Einstein equations (with the
gauge field arising from \KK\ reduction on an $S^1$), they
are also solutions, to leading order in $\alpha'$,
of low-energy string
theories
in less than ten dimensions when the compactification includes
an $S^1$ factor. In addition,
the Ramond-Ramond (RR) gauge field in type IIA string theory in ten dimensions
arises from dimensional reduction from eleven dimensions. We can thus
construct an instanton describing the nucleation of
a spherical six-brane carrying magnetic RR charge
in this theory in ten dimensions.

The third ingredient in our construction
is the observation that after dimensionally
reducing a  $D$ dimensional vacuum solution via a $U(1)$ isometry
to $(D-1)$ dimensions,
one can apply
a duality transformation which replaces the Maxwell two form
with a $(D-3)$-form field strength.
This yields electric analogs of the magnetic $p$-branes.
For the case $D=6$ ({\it i.e.} five reduced spacetime
dimensions) the resulting action is precisely the low energy
string effective action involving the metric, dilaton, and three-form $H$. The
duals of the magnetic strings  (one-branes) turn out to be fundamental strings.

The layout of the paper is as follows.
As we mentioned, section 2 contains an analysis of
the fixed points sets of general $U(1)$ isometries in arbitrary
dimensions.
We also construct the closely related
generalizations of the Melvin magnetic flux tube solution of
$D=5$ \KK\ theory. These are thickened branes of magnetic flux or
``fluxbranes," which are the appropriate backgrounds for nucleating
$p$-branes.
In section 3 we review various properties of the five-dimensional
Kaluza-Klein monopole, monopole
anti-monopole pairs, and pair creation.
In section 4 we present solutions describing spherical,
magnetically charged $p$-branes expanding in magnetic fluxbranes.
The Euclidean sections of these solutions are the instantons
for the nucleation of magnetically charged $p$-branes in
magnetic fluxbranes.
We also present solutions describing static magnetic $p$-branes
being held in unstable equilibrium by the fluxbrane.
In addition, we discuss related solitons that do
not carry magnetic charge. This is extended in section 5 to allow more
general, and more physical, values of the magnetic field at infinity
and we give the production rates for nucleating the charged branes.
In section 6 we show how dualizing the $D=6$ magnetic string
yields the fundamental string in five spacetime dimensions. Thus, the duals of
our magnetic string solutions describe a loop of fundamental string
in static unstable equilibrium in an electric field, and also a
loop that is expanding in an electric field. The latter, when
Euclideanised, is the instanton for the nucleation of a single loop
of fundamental string. We calculate the rate for this process.
Some concluding remarks are given in section 7. In the appendix we
describe in more detail the calculation of the instanton actions.

Since manifolds of many different dimensions will abound, we
will adhere to the convention that $D$ refers to the
dimension of a spacetime, $d$ to the dimension of
a Euclidean manifold that is to be considered as a spatial section
of spacetime,  and that these will
always refer to the dimensions of the unreduced geometry.

\newsec{Properties of higher dimensional symmetries}

In this section we will give
some general results which will be useful later. The theory
we start with is vacuum gravity in $D$ dimensions with action,
up to boundary terms, given by
\eqn\vacuum{S={1\over 16\pi G_D} \int d^Dx \sqrt{-g_D}{}\; R(g_D) .
}
If a geometry $ds_D^2$ has a Killing vector $\p /\p {x^D}$ with
closed orbits and
\eqn\red{ds_D^2 = e^{-{4\over\sqrt{D-2}}\phi}\(dx^D + 2 A_\mu dx^\mu\)^2
+ e^{{4 \over (D-3)\sqrt{D-2}}\phi} g_{\mu\nu} dx^\mu dx^\nu \ ,
}
then the action can be re-expressed as
\eqn\redact{S={1\over 16\pi G_{D-1}} \int d^{D-1}x \sqrt{-g}\[R(g) -
{4\over D-3}(\nabla \phi)^2 - e^{-4{\sqrt{D-2}\over D-3} \phi} F^2\]\ ,
}
where $ 2 \pi R G_{D-1} =  G_D$ and $R$ is the
radius of the compactified dimension.
The $D-1$ dimensional fields -- dilaton, $\phi$, gauge potential, $A_\mu$ and
metric, $g_{\mu\nu}$ -- can be read off from \red.

\subsec{Classification of fixed point sets}

If the isometry generated by the Killing vector above
has fixed points, then $\phi$ diverges and the metric $g_{\mu\nu}$
will be singular at those
points.
Let us consider
the general  classification of fixed
points of a $U(1)$ isometry in a $d$-dimensional Riemannian
manifold $M$. This is a straightforward generalization of the
four-dimensional case, which was analyzed in
\gibhaw. Let $\q$ be the associated Killing field, and consider the tensor
\eqn\tensorq{
q _{\alpha \beta} \equiv q_{\alpha ;
\beta}
}
at a fixed point $x$ where $\q =0$.
By virtue of Killing's equation $q _{\alpha \beta}$ is antisymmetric.
Let $V$ denote the kernel of $q _{\alpha \beta}$
and suppose dim $V = p$. Then vectors in $V$ are directions in the
tangent space at the fixed point, $T_x$, which are invariant under the
action of
the symmetry. Since the exponential map commutes with the symmetry action,
it follows that there is a $p$ dimensional subspace of fixed points. One
can show that this subspace is always totally geodesic. In four dimensions,
the only possibilities are $p=0$ and $p=2$: the first case is called
a `nut' and the second,
a `bolt'.
In higher dimensions, there are clearly more possibilities. Notice that since
the rank of a skew matrix must be even, there can  be no isolated fixed
points when $d$ is odd. In particular, ``There are no NUTS
in 11-dimensions" \refs{\ramond}.
In general, when $d$ is odd (even), $p$ is odd
(even).

The two form $q_{\alpha \beta}$
determines an element of the Lie algebra
$so(d)$ or, equivalently, a $U(1)$ subgroup of  $SO(d)$
that winds around a maximal torus.
The windings are determined
by the skew eigenvalues of $q_{\alpha \beta}$ in an orthonormal frame.
There are at most $[d/2]$ such eigenvalues, where $[r]$ denotes the
integer part of $r$. The
eigenvalues must all be rationally related and so determine up to $ [d/2]$
integers, $n_i$, some possibly zero, with  no common factor. These
integers can be viewed as the number of $2\pi$ rotations in different
orthogonal
two-planes in $T_x$ induced by one orbit of the isometry.

Near a fixed point, $M$ looks locally like ${\bf R}^d$ and we can analyse the
character of the different actions by identifying the space and the
tangent space $T_x$. Suppose the number of non-zero $n_i$ is $k$. Then
restricting to the $2k$ dimensions acted on by the rotation,
we can write the metric as
\eqn\many{
ds^2 = \sum_{i=1}^k \(d\rho_i^2 + \rho_i^2 d\vp_i^2\) ,
}
and
\eqn\manyq{\q = \sum_i n_i  {\p\over \p \vp_i.}
}
Introducing complex coordinates $\{Z^i\equiv \rho_i e^{i\vp_i}\}$
we can write the
circle action as the holomorphic action
\eqn\holo{  (Z^1,\dots Z^k) \rightarrow (
e^{i n_1 y} Z^1, \dots e^{i n_k y} Z^k),
}
where $y$, $0\le y < 2\pi$, parametrises the $U(1)$ subgroup.

The $U(1)$ subgroup acts freely away from
the isolated fixed point at $\rho_i=0$ $\forall i$. It follows that
topologically we have a principle $U(1)$
fibration of the odd-dimensional $(2k-1)$-sphere
given by
\eqn\sphere{\sum_{i=1}^k \rho^2_i = {\rm constant}.}
In the special case that all the $n_i=1$ this is the Hopf action giving
rise to the Hopf fibration
 \eqn\fibra{\matrix {S^1&{\buildrel i \over \longrightarrow}&S^{2k-1}\cr
            {}&{}&\;\;\bigg\downarrow \pi\cr
             {}&{}& CP^{k-1}\cr}}
For the case $k=2$ this is  the familiar --- from magnetic monopole theory ---
Hopf fibration of $S^3$ since
$CP^1 \cong S^2$.
The sign of $n_i$ can be changed by changing the sign of $\vp_i$; however
changing an odd number of signs changes the orientation of
space and gives the ``anti-Hopf'' action.

If the $n_i$ are not all equal
to $1$ or $-1$,
then $\q$ still acts without fixed points on $S^{2k-1}$. However, in this
case, the quotient metric will have conical singularities. This can be
illustrated by the simplest example, $k=2$:
\eqn\rfour{ds^2 = d\rho_1^2 + \rho_1^2 d\vp_1^2 +
d\rho_2^2 + \rho_2^2 d\vp_2^2\ .}
Consider the Killing vector
\eqn\killq{\q=n_1 {\p\over \p \vp_1} + n_2 {\p\over \p \vp_2},}
and let $\vp' = \vp_1$ and $\vp = \vp_2 - {n_1\over n_2}\vp_1$
which is constant along the orbits of $\q$. Then
$\q = n_1 {\p\over\p\vp'}$ and when we reduce, and also restrict the metric
to the $\rho_1^2 + \rho_2^2 = 1$ surface by setting $\rho_1=\cos\theta$
and $\rho_2=\sin\theta$, we obtain
\eqn\twod{ds_{II}^2 \propto d\theta^2 + {\sin^2\theta
\cos^2\theta
\over \cos^2 \t + {n_1^2\over n_2^2} \sin^2\theta }d\vp^2.
}
The range of $\theta$ is $0\le\theta\le \pi/2$ and the condition that
there be no conical singularities at $\theta=0, {\pi\over 2}$ is that
$\vp$ has period $2\pi=2\pi|{n_1\over n_2}|$. Since $n_1$ and
$n_2$ are co-prime, this condition cannot be satisfied unless $n_1^2=n_2^2=1$.

For a Lorentzian manifold, the
general classification of fixed point sets is more
complicated.
The main difficulty is that one cannot always
bring the generator of rotations  to block diagonal form.
Consider $so(2,1)$ for example. Any non-vanishing skew $3\times 3 $
matrix has a one-dimensional kernel.
The kernel may be timelike, spacelike or null. In the first two cases
one has a rotation or boost respectively. These cases admit
a block diagonal form with one block the $1 \times 1$ zero matrix
and the complementary block in the orthogonal two-plane is
a skew $2\times 2$ matrix.
If the kernel is timelike one has a
conventional axis of rotational symmetry.
If the kernel
is spacelike the fixed point set is locally like the
boost-invariant Boyer-axis of a black hole.
If the kernel is null however, corresponding to a so-called null rotation,
this reduction  cannot be done because there is no uniquely defined
orthogonal 2-plane.
However, it remains true that
even for Lorentzian metrics the fixed point sets will
be totally geodesic surfaces. Since these fixed points are often located at the
center of a soliton, it follows that the soliton obeys the
equations of motion of a ``fundamental" $p$-brane.
This will be true, in particular,
for all the $p$-branes discussed later.

\subsec{Fluxbranes}

It was shown in \refs{\dgkt,\dggh,\decay} that a uniform magnetic field in four
spacetime dimensions,
a generalization of the Melvin solution of Einstein-Maxwell
theory \refs{\melvin}, can be obtained by a dimensional reduction of a
five-dimensional geometry which is flat.
This five-dimensional spacetime
is obtained by starting with five-dimensional  Minkowski spacetime, $M^5$,
and identifying points under a combined spatial translation and rotation. When
the rotation is zero, one obtains the standard Kaluza-Klein vacuum.
When it is non-zero,
the field configuration in the reduced space (in which form it was
originally discovered \gm) is
that of an infinitely long straight
magnetic flux tube.
The generalization to magnetic ``fluxbranes"  in
higher dimensions is straightforward. Since time plays no role in the
construction, we start with $d$-dimensional Euclidean space. Higher
dimensional generalizations of the $d=4$
case are obtained by
identifying points under an element of the Euclidean
group which acts without fixed points on $\R^d$.
The question is simply to characterize
such elements.

The general element of the Lie algebra of the Euclidean group
$e(d)$ is a pair consisting of a translation
and a rotation with infinitesimal parameters given by
$v_\alpha$ and $\omega _{\alpha \beta}$,
respectively.
To obtain a nontrivial
gauge field, we need  $\omega _{\alpha \beta} \ne 0 $
and to avoid fixed points we require that the equation
\eqn\nofixed{w\cdot x + v=0,}
have no solution. This is the case if and only if
the kernel of $\omega$ is non-zero and $v$ has a component in the
kernel.
Under a change of origin in ${\bf R}^d$ by an amount $a_\alpha$,
$\omega $ is unchanged but $v$ changes to $\widetilde v$ given by
\eqn\change{
\widetilde v = v + \omega\cdot a.
}
One may always  choose $a$ so that the translation
$\widetilde v$ lies entirely in the kernel of $\omega$. This means that
the general Euclidean motion consists of a translation
combined with a rotation in an orthogonal hyperplane.

Since the translation part of the symmetry just fixes the scale
of the internal
direction, it follows that the classification of these magnetic fields
reduces to the classification of rotations in ${\bf R}^{d-1}$. A
rotation is given by a two-form with $[(d-1)/2]$ skew eigenvalues
so a magnetic field configuration is specified by $[(d-1)/2]$ real
numbers, $B_i$.

We can ask what these field configurations look like in the
reduced spacetime. Consider $d=2m+1$ (the $d$=even case is similar).
\eqn\mag{
ds^2 = \sum_{i=1}^m \(d\rho_i^2 + \rho_i^2 d\vp_i^2\) + dy^2\ .
}
We identify each point  with the point obtained by moving a distance $2\pi R$
along the integral curves of the Killing field
\eqn\bigq
{\q =  {\p\over \p y} +  \sum_{i} B_i{\p\over\p\vp_i} \ .
}
Introducing the new coordinates  $\tvp_i=\vp_i -B_i y$ which are
constant along
the orbits of $\q$,  we find that
$\q=  {\p\over\p y}$ and the above identification just consists of making $y$
periodic with period $2\pi R$ at fixed $\tvp_i$.
The flat metric \mag\ now takes the form
\eqn\fluxbrane{
ds^2 = \Lambda\[dy + {1\over \Lambda}
\sum_i B_i\rho_i^2 d\tvp_i\]^2 +
\sum_i \( d\rho_i^2 + \rho_i^2 d\tvp_i^2\) - \Lambda^{-1}\(\sum_j B_j \rho_j^2
d\tvp_j\)^2\ ,
}
where
\eqn\dila{ \Lambda= 1 + \sum_i B_i^2 \rho_i^2 \ .}
The reduced metric, dilaton and gauge potential can be read off from
\fluxbrane\ (after adding an extra
time direction) using
\red\ with $D=d+1$ to give
\eqn\redmet{
\eqalign{ds^2_{D-1}=& \Lambda^{1\over D-3}\[-dt^2+
\sum_i \( d\rho_i^2 + \rho_i^2 d\tvp_i^2\) - \Lambda^{-1}\(\sum_j B_j \rho_j^2
d\tvp_j\)^2\]\ , \cr
e^{-{4\over \sqrt{D-2}}\phi} =& \Lambda, \qquad
A= {1\over 2\Lambda} \sum_i B_i\rho_i^2 d\tvp_i \ .\cr
}}

When only one $B_i$ is non-zero this is a
thickened brane of magnetic flux, or ``fluxbrane,'' of dimension $(d-3)$ and
was found by Gibbons and Maeda \gm.
The amount of  flux  passing through a
one-dimensional loop,  $\gamma$,
is given by integrating the one-form potential, $A$,
 around the loop  (note
that in $(d-1)$ spatial dimensions a circle surrounds a $(d-3)$-brane):
flux $=\int_{\gamma} A$.
Each non-zero parameter $B_i$  adds another orthogonal
fluxbrane to the configuration. The gauge field strength is
maximised at the intersection of the centers of the
fluxbranes, which is the fixed point set of the rotational part
of $\q$. When
$k$ of the parameters are non-zero this is a $(d-2k-1)$-hyperplane.
The generic configuration in the $(d-1)$-dimensional reduced space is a set
of $[(d-1)/2]$ orthogonal $(d-3)$-fluxbranes.
These intersect in a point when $d=$odd and in a line when $d=$even.

\newsec{Five-dimensional monopoles}

In this section we
shall review some of the geometrical and topological properties of the
basic Kaluza-Klein monopole \refs{\rafkk, \grossperry}
relating them to the fixed point set
analysis of the previous section. We also present a novel interpretation of
the four-dimensional Euclidean Schwarzschild solution.

\subsec{The basic monopole}

The single monopole  solution is a five-dimensional spacetime which is
a metric product ${\bf R} \times M$ of a time factor with coordinate $t$ and
Euclidean Taub-NUT space, $M$. $M$ is Ricci flat, self-dual, topologically
$\R^4$, and admits an
isometric circle action. The metric is, explicitly,
\eqn\mon{
 ds^2 = -dt^2 + V^{-1} (d y + 2A_\vp d\vp)^2
  + V (dr^2 + r^2 d\Omega)
}
with
\eqn\ang{ A_\vp = 2m(1-\cos\t), \quad \quad  V = 1 +  { 4m \over r}\ .}
The period of $y$ is $2\pi R$ with $R=8m$.
The Killing vector associated with the $U(1)$ isometry
is ${\bf q} = \p/ \p {y}$.
If the circle action was free, the topology
of $M$ would be that of a,  possibly twisted, circle bundle
\eqn\fib{
\matrix {S^1&\longrightarrow&M\cr
           {}&{}&\;\;\big\downarrow \cr
           {}&{}& \Sigma\cr}}
over some     complete non-singular 3-manifold
$\Sigma$. One could then think of the $S^1$ factor globally as an internal
manifold. Because of the fixed point $r=0$ at which $g_{yy}$
vanishes, and hence the length of the circle fibres goes to zero, this
description is only valid  away from $r=0$.

At the center of the monopole the manifold is smooth and locally
indistinguishable from the flat metric on ${ \bf R}^4$.
As discussed in the previous section, at a fixed point, the
circle action may be thought of as a rotation in two orthogonal
two-planes in ${ \bf R}^4$, characterised by two integers, $n_1$ and $n_2$.
For a single \KK\ monopole we have  $n_1=n_2=1$, {\it i.e.} the
Hopf action on small $S^3$'s surrounding the center.
The reduced four-dimensional
spacetime is thus free of singularities except at the center of the monopole.

Antimonopole solutions are given by \mon\
with the opposite sign of $A_\vp$. Now the $U(1)$ action
near the antipole center is labeled by $n_1=-n_2=1$: it is the
anti-Hopf action.

\subsec{Static monopole-antimonopole pairs}

It is interesting to ask what the topology of a \mam\ configuration
would be. Physically one would not expect a static
asymptotically vacuum solution since the pair will attract,
so one must  either give up the asymptotic
vacuum condition  or suspend the field equations. In both cases the topology
should be the same.   We will argue that the
topology is in fact $\R^2 \times S^2$. Let the spatial 4-manifold be $M$, then
$M = A\cup B\cup C$
where
 $A$ and $B$ are both  four-balls $D^4$ corresponding to the monopole
and antimonopole and
$C$ is the non-trivial $U(1)$ bundle over $\R^3\#D^3\#D^3$ ($\R^3$ with
two three-balls removed)
which has zero winding over the sphere at infinity, and windings $+1$
and $-1$ over the other two $S^2$ boundaries \refs{\sorkin}.
Since the bundle is trivial over the sphere at infinity we can
add in an $S^1$ there:
$ M \cup S^1  = A \cup B \cup C'$ where $C'$ is a U(1) bundle over
$S^3 \#D^3\#D^3=S^2\times D^1$ where the one-ball, $D^1$, is just
the one-dimensional interval.
The U(1) must have unit twist over one $S^2$
boundary and unit anti-twist over the other $S^2$ boundary.
So $C' = S^3 \times D^1$
which is $C' = S^4 \#D^4\#D^4$. So $M \cup S^1 = S^4$ and
$M = S^4 - S^1 = \R^2 \times S^2$.

Thus, we see that the
monopole-antimonopole manifold has the same  topology  as the
four-dimensional   Euclidean Schwarzschild solution:
\eqn\schw{
 ds^2 = \(1-{ 2m\over r} \)
 d \tau^2 +\(1-{ 2m\over r} \)^{-1} dr^2  + r ^2 (d \theta ^2
+\sin^2 \theta d \vp ^2 )     \ ,
}
where $\tau$ has period $2\pi R$, $R=4m$.
Indeed we will see that in a certain precise
sense \schw\ can be regarded as containing a
\mam\ pair.
Firstly, if we add on a trivial time direction, we can regard
\schw\ as a static five-dimensional solution
describing a minimal two-sphere in space, a
``bubble,'' poised in
unstable equilibrium. If we reduce this five-dimensional
solution to four dimensions along the orbits of the Killing
vector ${\partial\over\partial\tau}$ we find that the
minimal two-sphere or bolt, being a fixed point set of that
circle action, looks singular in four dimensions.

Consider now the alternative Killing field
\eqn\killing{ {\bf q} =
{\partial \over \partial \tau} + {1\over R} {\partial \over \partial \vp}\ .
}
 It has
fixed points at the north and south poles, $\theta = 0, \pi$ of the
2-sphere  $r= 2m$. It can be shown that
at one, the action is the Hopf action,
and at the other it is the anti-Hopf action.
At each pole, the two orthogonal planes in which the rotations act are the
tangent spaces to the
$(r, \tau)$-section at the horizon (``tip of the cigar'') and
the $(\theta, \vp)$ horizon 2-sphere.
Near infinity, however,
${\bf q}$  becomes a linear combination of a {\it translation} and a rotation.
As discussed in section 2.2, a magnetic field in \KK\ theory is obtained
by taking the quotient of flat space with respect
to precisely this type of symmetry.
If we therefore take the closed orbits of ${\bf q}$
as our internal \KK\ circles, we can interpret \schw\ in four dimensions
as a static monopole-antimonopole pair held apart
by a background magnetic field with $B=1/R$.
One would expect such a configuration to be unstable and it is: the
negative mode of the
four-dimensional Euclidean Schwarzschild solution
gives rise to an exponentially growing mode of this five-dimensional solution.

\subsec{Dynamical \mam\ pairs and pair creation}

A key observation of  \refs{\decay} was that one may relate the
{\it{five}}-dimensional
Schwarzschild solution to monopole-antimonopole
pair-production.\foot{To obtain arbitrary values of the magnetic field at
infinity, one should consider the five-dimensional Kerr solution. We will
return to this in section 5, but for now, we illustrate the construction using
the simpler Schwarzschild solution.} Consider the
five-dimensional Euclidean Schwarzschild solution
which we write as
\eqn\bubble{
 ds^2 = \[ 1- \({ \r \over r}\)^2  \] d  \tau^2 +
 \[ 1- \({ \r \over r}\)^2  \]^{-1}dr^2   + r^2 \[ d \alpha ^2 +
\cos ^2 \alpha ( d \theta ^2 + \sin ^2 \theta d \vp ^2 )\],
}
where $\tau$ has period $2\pi R$ with $R=r_H$, and $-\pi/2 \le
\alpha \le \pi/2$.
{} From the purely five-dimensional point of view, this is an
instanton that describes the decay of $M^4 \times S^1$ where
$M^4$ is four-dimensional Minkowski space. It has topology
$\R^2 \times S^3$ and it asymptotically
approaches flat  $\R^4 \times S^1$. It has a zero momentum slice
$\alpha= 0$ which has topology
$\R^2\times S^2$ and contains a minimal
two-sphere or ``bubble,'' $r=\r$: overall, the zero-momentum
slice is very similar
to four-dimensional Euclidean Schwarzschild. The subsequent Lorentzian
post-decay evolution is obtained by setting $\alpha = i t$
in \bubble\ and describes the minimal two-sphere
expanding. Its area increases like $\cosh^2 t$ and this
solution looks like a dynamical version of the static
bubble of the previous subsection.

If one reduces along $\partial \over \partial \tau $ one may think of this
as  an instanton for the decay of the
vacuum in four dimensions \refs{\witten}.
The fixed point set restricted to the zero-momentum slice
is the entire minimal two-sphere or bubble which appears
singular in the reduced spacetime.
However, one may  alternatively reduce along ${\bf q}$ with $R=\r$
in \killing.
In four dimensions, in this case, the instanton describes the decay of
a magnetic field via pair creation of a \mam\ pair. We can
see this by
subjecting the  zero momentum
slice to the fixed point set analysis of the previous
subsection. The pole and antipole are the fixed points of the
circle action of ${\bf q}$ and the twisting character of
${\bf q}$ at infinity means the particles are immersed in a
magnetic field of strength $B=1/R$. Now, however, in the subsequent Lorentzian
evolution the particles do not stay in
their static positions but accelerate apart.

\newsec{Higher dimensional generalizations}
\subsec{Flat charged $p$-branes}

The simplest generalization of the basic \KK\ monopole
is to take the product of \mon\ with an arbitrary number, $p$,
of flat directions before doing the \KK\ reduction. This gives
a magnetically charged $p$-brane in $(p+4)$ spacetime dimensions,
e.g., a magnetically charged string in five dimensions.
If the
extra dimensions are infinite, the $p$-brane is also  infinite.
(We could, of course, consider the extra dimensions to be a torus
in which case the brane is also a torus, but this would change the topology
of the
reduced spacetime. Moreover if the torus was large this solution would
approach the infinite brane and if it were small it would reduce,
in a Kaluza-Klein sense, to the monopole again.)

An obvious instanton describing the production of a pair of these
$p$-branes in a magnetic field
is obtained by taking the product of five-dimensional
Euclidean Schwarzschild with the extra flat dimensions
and reducing via \killing.
The asymptotic magnetic field configuration of this
instanton is a $(p+1)$-dimensional fluxbrane.
However, if the
extra dimensions are infinite then the action for this instanton
is infinite, even relative to the background magnetic field.

\subsec{Spherical charged $p$-branes }

It might appear that the $(p+1)$-dimensional fluxbrane cannot
decay because the instanton that describes pair creation of
infinite magnetic $p$-branes has infinite action. This is,
in fact, not the case. We will see in this
section that it can decay by the nucleation of
a single $p$-brane with topology $S^p$.

We start with $p=1$ as an example, and first
describe a magnetically charged loop of string in static
equilibrium in a background magnetic field.
As discussed in section 2, a magnetic
field can be obtained by taking six-dimensional Minkowski spacetime,
\eqn\sixmink{ ds^2 = -dt^2 + d\tau^2 + dr^2 + r^2 d\vp^2 + dx_i dx^i\ ,}
and identifying points by moving a distance $2\pi R$ along the
Killing vector
\eqn\fdsym{ \q = {\p \over \p \tau} + {1\over R} {\p \over \p \vp}.}
The reduced spacetime
describes a two-dimensional magnetic fluxbrane with $B=1/R$.

To obtain a charged loop of string in this background, we start with
the product of time and the five-dimensional Euclidean
Schwarzschild solution
\eqn\magstr{
ds^2 = -dt^2 + \[ 1- \({ \r \over r}\)^2  \] d  \tau^2 +
\[ 1- \({ \r \over r}\)^2  \]^{-1} dr^2   + r^2 d\Omega_3\ .}
The metric on the three-spheres can be
written\foot{To see this, start with \many\ and \sphere\ with $k=2$. Then
set $\rho_1 = \sin\theta,\ \rho_2 = \cos\theta$.}
\eqn\thrsp{ d\Omega_3 = d\theta^2 + \sin^2\theta d\vp^2 +
\cos^2 \theta d\chi^2 \ ,}
where $0\le \theta \le \pi/2, \ 0\le \vp, \chi \le 2\pi$. We now reduce
down to five dimensions using the symmetry \fdsym.
The fixed points of this Killing field are at $r=\r, \ \theta = 0$ which
is clearly a circle parametrized by $\chi$. For each value of $\chi$,
the indices of the symmetry at the fixed point (as discussed in section 2)
are the same as the usual Kaluza-Klein
monopole. This shows that in the reduced spacetime
there is a  magnetic charge at each fixed point on the string.
The solution thus describes a circular charged string.
Asymptotically, the solution approaches the two-dimensional fluxbrane.

To be explicit, we set $\tvp = \vp - (\tau/\r)$ in \magstr\ (so $\tvp$ is
constant along the orbits of the symmetry \fdsym). We then compare
the resulting metric with \red\ setting $D=6$ and $x^D = \tau$. The
result is
\eqn\restsn{
\eqalign{
e^{-2\phi} &= 1 - \({ \r \over r}\)^2 + \({ r \over \r}\)^2 \sin^2
	    \theta, \qquad A = {r^2 \sin^2\theta \over 2\r} e^{2\phi} d\tvp,\cr
ds^2 &= e^{-2\phi/3}\( -dt^2 + \[1 - \({ \r \over r}\)^2\]^{-1}dr^2
 +r^2(d\theta^2 + \cos^2 \theta d\chi^2) \)}}
 $$+ e^{4\phi/3} \sin^2\theta (r^2 - \r^2)
 d\tvp^2 \ .$$
Near $r=\r$, the metric takes the form
\eqn\redstr{ ds^2 \approx e^{-2\phi/3} \[ -dt^2 + d\tilde r^2
  + \r^2( d\theta^2 + \cos^2\theta d\chi^2)\]
  + e^{4\phi/3} \tilde r^2\sin^2\t d\tvp^2 \ ,}
where $\tilde r^2 = r^2 - \r^2$ and
\eqn\phinrh{ e^{-2\phi} \approx \sin^2\t +
  \({\tr \over \r}\)^2 \(1+ \sin^2\t\) \ .}
The singularity
$\tr =0, \ \t =0$ is a ring representing the loop of string.
Notice that $\tr =0, \ 0< \t \le \pi/2$ is a regular
two-dimensional surface spanning the ring singularity.

We now discuss the nucleation of a
closed magnetically charged string in our two-dimensional
magnetic fluxbrane.  The appropriate instanton
is the six-dimensional Euclidean Schwarzschild solution
\eqn\schsd{
 ds^2 = \[ 1- \({ \r \over r}\)^3  \] d  \tau^2 +
 \[ 1- \({ \r \over r}\)^3  \]^{-1} dr^2   + r^2 (d\alpha^2 + \cos^2 \alpha
 d\Omega_3) \ ,}
where $\tau$ has period $2\pi R$ with $R= 2\r/3$, and $-\pi/2 \le \alpha \le
\pi/2$.
By direct analogy with the five-dimensional instanton discussed in the
previous section on the basic monopole,
this instanton has a zero momentum surface
$\alpha = 0$ which contains a minimal
three-sphere or ``bubble.''
The subsequent Lorentzian evolution is obtained by setting
$\alpha = it$. The minimal three-sphere  expands exponentially in $t$.
Reducing \schsd\ along ${\p\over\p\tau}$ we obtain an instanton
describing vacuum decay in this $D=6$ \KK\ theory.
If we reduce \schsd\ along the symmetry \fdsym\ we obtain an
instanton which is asymptotically a magnetic fluxbrane
of strength $B=1/R$. Restricted to the
surface $\alpha = 0$, the reduction is virtually identical to the one
described in the static case
above. In particular, one has a charged loop of string. In the subsequent
Lorentzian evolution the loop expands since it lies on the  expanding
bubble.

For the Schwarzschild instanton, the single parameter $R$ (or $\r$)
governs both the strength of the asymptotic
magnetic field and the charge on the string. We will
see in the next section that this value of the magnetic field,
$B=1/R$, is unphysically large
and we will construct solutions corresponding
to the decay of more physical values of $B$ by considering instantons
based on higher dimensional rotating black hole solutions. We will also
calculate the rate of nucleation in the semi-classical approximation.

One might wonder why the nucleation of a charged loop of string
does not violate charge conservation.
The point is that in four (reduced)
spatial dimensions, the total magnetic charge
of any localized object must be zero. This is because the magnetic charge
is obtained by integrating $F_\mn$ over a two-sphere. One cannot integrate
$F_\mn$ over the three-sphere at infinity. This does not contradict the
fact that locally, the string carries a magnetic charge. The magnitude of
the charge $q$ at the fixed points is $R/4$, where $R$ is determined by the
periodicity in the compact direction. However, the sign of the charge depends
on
a choice of orientation. If one charge is
chosen to be $+q$, the opposite one is necessarily $-q$.  This follows
from the fact that the orientation induced on the two-sphere enclosing
a point of the string depends on the tangent vector to the string, which
points in the opposite direction halfway around the loop.
If one takes a
slice through the Lorentzian solution, it describes oppositely charged
monopoles accelerating apart in a magnetic field.
More generally, any $p$-brane,
$p\ne 0$, which locally carries a charge associated with
any
$r$-form $F_r$, must have zero net charge when the $p$-brane is confined to a
compact region. The reason is simply that the sphere at infinity will have
dimension $(p+r)$ which is larger than the rank of $F_r$.

The construction of magnetically
charged spherical $p$-branes for $p>1$ is a straightforward
extension of these ideas to higher dimensions. Taking the quotient of
$(p+5)$-dimensional Minkowski spacetime by the symmetry \fdsym\
yields a $(p+1)$-dimensional magnetic fluxbrane.
This can support  a static charged spherical $p$-brane as follows.
Consider the Euclidean Schwarzschild solution in $d=(p+4)$ dimensions
crossed with a trivial time direction
 \eqn\pdsch{
 ds^2 = -dt^2+\[ 1- \({ \r \over r}\)^{p+1}  \] d  \tau^2 +
 \[ 1- \({ \r \over r}\)^{p+1}\]^{-1} dr^2   + r^2 d\Omega_{p+2}\ ,}
where $\tau$ is a periodic coordinate with period $2\pi R$
and $R=2 r_H/(p+1)$.
The metric on the $(p+2)$-sphere can be written
\eqn\pdsph{d\Omega_{p+2}=
d\theta^2+\sin^2\theta d\vp^2+\cos^2\theta d\Omega_{p}\ ,
}
with $0\le\theta\le\pi/2$.
We now
reduce along
the symmetry  \fdsym. The fixed points again occur at $r=r_H$ and
$\theta=0$ so that on a static slice,
the set of fixed points is now a sphere
$S^p$.
At each fixed point, the behavior of the symmetry in the four
directions orthogonal to the $S^p$ is again exactly that of a Kaluza-Klein
monopole. So we obtain
a static charged spherical $p$-brane.

To nucleate such a $p$-brane, we start with the $D=(p+5)$-dimensional Euclidean
Schwarzschild solution,
\eqn\nucleate{ds^2 = \[ 1- \({ \r \over r}\)^{p+2}  \] d  \tau^2 +
 \[ 1- \({ \r \over r}\)^{p+2}\]^{-1} dr^2   + r^2\(d\alpha^2
+\cos^2\alpha d\Omega_{p+2}\)\ ,
}
with $d\Omega_{p+2}$ given by \pdsph. Reducing this
via \fdsym\ we obtain an instanton for the nucleation of
the $p$-brane. The subsequent Lorentzian evolution is
 described by the  analytic continuation
$\alpha=it$ and
in this Lorentzian spacetime the
fixed point set of \fdsym\  is the world volume of a spherical
charged $p$-brane that exponentially expands.

There are many instantons which asymptotically
approach the same magnetic field. One can start with the product
of a $(p+5-m)$-dimensional Schwarzschild solution and $\R^m$. To maintain
the same asymptotic magnetic field, we always reduce under \fdsym.
These instantons describe the nucleation of charged
$p$-branes  with topology $\R^m \times S^{p-m}$. However, the action for all
of these instantons is infinite due to the infinite volume of $\R^m$.

An interesting application of this construction is
to the type IIA string. The low energy action of this theory in ten
dimensions contains a Ramond-Ramond gauge field which comes from
Kaluza-Klein reduction of an eleven-dimensional metric.
The \KK\ six-brane we have been considering thus carries
RR charge \townsend.
Starting with
the $D=11$ Minkowski space, and taking the quotient under \fdsym\
one obtains a seven-dimensional magnetic fluxbrane.
This is unstable to the nucleation
of a spherical, magnetically charged six-brane. The appropriate
instanton is simply the $D=11$ Schwarzschild solution.

\subsec{Uncharged $p$-branes}

One can also construct static $p$-branes in $d-1$ reduced
spatial dimensions where $p+4<d$. These do not
carry magnetic charge, but they can arise when fluxbranes intersect.
As described in section 2, intersecting fluxbranes
are obtained by taking the quotient of  Minkowski spacetime
under a symmetry which is a translation plus a rotation,
where the rotation is not restricted
to lie in a single two-plane. We first consider static solitons
and then discuss how one can nucleate such objects.

We begin with the product of time and the
$d$-dimensional Euclidean Schwarzschild metric. To describe $k$
orthogonal fluxbranes asymptotically, we write the metric on the
$(d-2)$-spheres
in terms of $\theta_i, \vp_i, \ i=1,...,k$, and coordinates on a
$(d-2-2k)$-sphere
by iterating \pdsph\ $k$ times. We then reduce along
the symmetry
\eqn\gensym{ \q = {\p \over \p\tau} + {1\over R}\sum_{i=1}^k {\p \over \p\vp_i}
\ ,}
where $\tau$ has period $2\pi R$.
Asymptotically, the solution resembles
\redmet\ with $k$ non-zero $B_i$'s and describes
$k$ orthogonal fluxbranes. Recall that in $(d-1)$ reduced  spatial
dimensions, each fluxbrane has dimension $(d-3)$ and $k$ of them will
intersect in a surface of dimension $(d-1-2k)$. The vector $\q$
has fixed points at $r=\r$ and $\theta_i=0$ for all $i$. This is a
spherical  $(d-2-2k)$-brane
which lies in the intersection of the fluxbranes.
For $k\ne 1$, this brane does not carry magnetic charge in the
reduced spacetime since it does not have the right dimension. Nevertheless,
it is a static, though unstable, localized excitation of the fields.

The configuration of $k$ intersecting fluxbranes is unstable to the
nucleation of such uncharged $(d-2-2k)$-branes. The instanton is the $(d+1)$
dimensional Euclidean
Schwarzschild solution and the  analytic
continuation is identical to that for the
case of the charged branes. The only difference is that one now reduces
along the more general symmetry \gensym.

If $k=1$, this instanton
construction reduces, of course, to the one in section 4.2.
There are
two other special values of $k$ of particular interest. If $d$
is even and $k= (d-2)/2$ (the maximal number) then the fixed point set
consists of two points. On a small sphere surrounding each point, the
symmetry acts
like the Hopf fibration \fibra, of $S^{d-1}$ by $S^1$.
Bais and
Batenberg \baisbat\ have constructed a Ricci-flat space containing a
single object of this type. It can be viewed as a generalization of
the Taub-NUT solution to higher (even) dimensions.
Topologically the manifold is simply ${\bf
R}^{d}$.  The metrics  admit a circle action, with a single fixed point.
Although the length of
the circle orbits tends to a constant near infinity, the Bais-Batenberg
solutions  for $d>4$
cannot be  regarded as circle bundles over  ${\bf R}^{d-1}$ asymptotically.
This is because the angular part of the reduced space is not $S^{d-2}$,
but $CP^{(d-2)/2}$. Thus, one could argue that these objects cannot
exist in isolation. However, we have seen that they can exist in pairs
and can, in fact, be pair created.

The second special case occurs when $d$ is odd and $k$ is again
maximal $k = (d-1)/2$. In this case, the symmetry \gensym\ has {\it no} fixed
points. It corresponds to a combination of
${\p\over\p\tau}$ and a Hopf rotation of the
$(d-2)$-sphere. The reduction then leads to a non-singular solution of the
Einstein-Maxwell-dilaton equations arising from the action \redact.

\newsec{Different asymptotic magnetic field values}

The solutions we have constructed from the Schwarzschild metric
have values of the
magnetic fields at infinity which are fixed completely by the
radius of the compactified dimension, $B=1/R$.
We should point out that this value of $B$ is actually
unphysically large in the following sense. Consider the single
fluxbrane solution with parameter $B$. The radius of the compactified
direction is not constant but grows from $R$ to infinity as the distance from
the center
increases. In order for the \KK\ reduction to make sense and
also for the configuration to be a reasonable approximation
of a uniform magnetic
field we should only consider distances from the center of the
fluxbrane $\rho< 1/B$.  This restriction occurs not only in
\KK\ theory, but also when considering pair
creation of black holes in magnetic fields in
Einstein-Maxwell theory in four dimensions.
However, in \KK\ theory we also have the condition that only
distances large with respect to the radius of compactification
should be considered in the reduced spacetime. Thus, $\rho>>R$ and
so we have the condition $BR<<1$.

The value of the asymptotic magnetic fields we have been considering
so far is well outside this physical range of validity.
We have allowed them up till now for reasons of simplicity:
the construction of the solutions describing the spherical charged $p$-branes
using the Schwarzschild metrics is simpler, though qualitatively the same as
the construction we will now give
of solutions describing the same types of branes but with arbitrary
values of the background magnetic field. We wanted to describe
the construction in a simpler setting before giving the
more complicated solutions that are of
more physical interest.
These are given by $U(1)$ reductions of
higher dimensional {\it rotating} black hole solutions.

Myers and Perry found analogues of the Lorentzian Kerr solution
for arbitrary spacetime dimension $N$
\refs{\mp}.  A particular case in $N=8$
dimensions
has been obtained independently by Chakrabarti \chakra\ using the special
properties of the octonions. As noted by Myers and Perry, a rotating body
in $(N-1)$ spatial dimensions has an associated angular
momentum which may be thought of as a 2-form $\omega$. Thus,
there are $[{N-1 \over 2}]$ rotation parameters $a_i$. These,
together with the mass, characterize the solutions.
If all the parameters $a_i$ vanish then one obtains the
usual higher-dimensional Schwarzschild solution.
One expects that, just as in four spacetime dimensions,
 the solutions are unique but to our knowledge there is no proof.
The general solution has continuous isometry group ${ \bf R} \times
SO(2)^{[{N-1 \over 2}]}$ but as more of the parameters become zero
the isometry group is enhanced, becoming ${ \bf R} \times
SO(N-1)$ in the non-rotating Schwarzschild limit. The case obtained by
Chakrabarti has $a_1=a_2=a_3$  and
is an example of slightly enhanced symmetry (there are
extra discrete symmetries).
In the general case one therefore has $[{N-1 \over 2}]$ ignorable azimuthal
coordinates $\vp_i$ parametrizing the maximal torus of $SO(N-1)$ and
which may be thought of near infinity  as rotations in $[{N-1 \over 2}]$
orthogonal 2-planes in ${ \bf R}^{N-1}$.

The Euclidean solutions,
for which $a_i=i\alpha _i$ with $\alpha _i$ real, are complete and
non-singular provided that a suitable periodic identification is
made. If we denote the Euclidean time by
$\tau =i t$, then one identifies points by moving a
distance $2\pi R$ along the integral curves of the Killing field
\eqn\idents{
\q=  {\p\over \p \tau} +  \sum_i \Omega_i {\p\over\p \vp_i}\ ,
}
where $R =  1 / \kappa $ and $\kappa $ and $i\Omega_i$
are the surface gravity and angular velocities respectively. General
expressions for
them may be found in \mp.

To construct solutions describing
spherical charged $p$-branes, it suffices to
consider the metrics with only one angular momentum parameter non-zero.
The $N$-dimensional Euclidean metric then takes the form
\eqn\eclkerr{ ds^2 = \(1- {\mu\over r^{N-5}\S}\) d\tau^2 -{2\mu \alpha
 \sin^2\t \over
r^{N-5}\S} d\tau d\vp  + {\S\over r^2 - \a^2-\mu r^{5-N}} dr^2 + \S d\t^2}
$$+{\sin^2 \t\over \S} [ (r^2 - \a^2) \S - \mu r^{5-N} \a^2 \sin^2 \t] d\vp^2
+r^2 \cos^2\t d\Omega_{N-4} \ , $$
where $\S = r^2 - \a^2\cos^2\t$ and $0\le\theta\le\pi/2$.
The horizon is located at $r = \r$ where
\eqn\kerrhor{ \r^2 = \a^2 + {\mu\over \r^{N-5}} \ . }
The radius of the circle at infinity is
\eqn\radinf{R = {1\over \kappa} = {2\mu\r^{6-N} \over (N-3)\r^2 - (N-5) \a^2}
\ ,}
while the Euclidean angular velocity is
\eqn\angvel{ \Omega = {\a \r^{N-5}\over \mu} \ .}
Their product is thus
\eqn\product{ \Omega R = {2\a\r\over (N-3)\r^2 - (N-5) \a^2}.}
This clearly vanishes when $\alpha = 0$. Now consider the limit
$\a \to \infty$. Considering \kerrhor\ and \radinf\ we find
that to keep $R$ fixed we need  $\r \to \a$ and
$ \mu \to R\a^{N-4}$.
Then \product\ implies that $\Omega R$ approaches one.
Similarly, the limit $\a \to -\infty$ sends $\Omega R\to -1$.
Thus, $\Omega R$ takes values between one and minus one.

To obtain the unstable,
static magnetically charged $p$-branes one starts with the
product of time and \eclkerr\ with $N=p+4$.\foot{
The case of the $N=4$ Euclidean Kerr solution with a flat time direction
added was considered in \refs{\grossperry}. There, the solution was
interpreted as a dipole but the presence of the background magnetic
field was unnoticed.}
If one reduces along
\eqn\oneparam{ \q = {\p\over\p \tau} + \Omega {\p\over\p\vp}}
the fixed point set will be the entire horizon
and we will obtain an unstable bubble
immersed in  a magnetic fluxbrane
of strength $B=\Omega$.
If instead, we reduce along
\eqn\qprime{ \q' = \q - {\sigma\over R} {\p\over\p\vp}\ ,}
where $\sigma=\Omega /|\Omega|$,  we obtain
the charged $p$-brane. The asymptotic value of the magnetic
field is $B= \Omega -(\sigma/R)$. $|B|$
can be made as small
as we like by tuning $\Omega$.
Note that as $B\to 0$,  $\r \to \infty$ and the `size' of the
$p$-brane becomes larger.
The limiting solution with $B=0$ is just the
noncompact $p$-brane obtained by taking the product of the standard
\KK\ monopole and $\R^p$.

We can clearly turn on additional fluxbranes at infinity by starting with
the black hole with several rotation parameters non-zero.
If we reduce using
\idents, the fixed point set is the horizon itself and we will
obtain an unstable bubble  living at the intersection of several fluxbranes
with $B_i= \Omega_i$. If we reduce using
$\q' = \q - (\sigma_j /R) {\p/\p\vp_j}$ (with $\q$ as in
\idents, $\sigma_j= \Omega_j /|\Omega_j|$, and no sum on $j$),
for any choice of $\vp_j$,
then we obtain
a magnetically charged spherical $p$-brane
in a background of intersecting fluxbranes
with $B_i= \Omega_i$ $i\ne j$, and $B_j = \Omega_j-(\sigma_j
/R)$.
Adding additional rotations to $\q'$ reduces the dimension of the
fixed point set which becomes an uncharged brane.

An instanton describing the nucleation of a charged $p$-brane
can be obtained from
\eclkerr\ with $N=p+5$ by reducing along
$\q'$ in \qprime.
This instanton corresponds to the nucleation of a spherical charged
$p$-brane in a background $(p+1)$-fluxbrane of strength $B=\Omega-\sigma/R$.
The Lorentzian solution, representing the post-tunneling evolution,
is obtained by analytically continuing in one of the ignorable
angles in $d\Omega_{N-4}$.
This appears to give a static solution rather
than the expanding solution we obtained earlier from the Schwarzschild metric.
However, the resulting
timelike Killing field is really a boost, and the spacetime
one obtains from Kerr is qualitatively similar to the one obtained from
Schwarzschild and describes the spherical $p$-brane expanding.
(For a detailed discussion of this in the  case of five
dimensions, see \decay.)

In the semi-classical approximation, the rate of nucleation is
given by $e^{-I}$ where $I$ is the Euclidean action of the
instanton. The action for the instanton
with one angular momentum parameter non-zero
is computed in the Appendix and is
\eqn\kerract{ I = {V_{p+1} \over 8(p+2)G_{p+4}}\mu,}
where $V_{p+1}$ is the volume of a unit $(p+1)$-sphere.
One can rewrite this in terms of the
magnetic field at infinity and the compactification radius, but the
expression is complicated and not very illuminating (see Appendix).
However, in
the limit where the asymptotic magnetic field $|B|$ is small, one finds
\eqn\weakmu{ \mu = \( {p+1\over 2}\)^{p+1} {R\over |B|^{p+1}}\ .}
The compactification radius $R$ is related to the charge on the $p$-brane
by the usual expression for \KK\ monopoles $q=R/4$, and this charge is
in turn proportional to the mass per unit $p$-volume or tension of the
$p$-brane. Thus, the nucleation rate $e^{-I}$ is increased by either increasing
$|B|$ or decreasing the tension, as expected.
For $p=0$, \kerract\ and \weakmu\ reduce to the Schwinger result
for pair creating
monopoles in a weak magnetic field \refs{\dggh}.

We close this section by noting that instantons describing the nucleation of
a spherical charged $p$-brane
in intersecting fluxbrane backgrounds can also be obtained. One
considers the Kerr solution with several non-zero rotation parameters
and reduces along $\q'= \q-  (\sigma_j/R) {\p/\p\vp_j}$
(with $\q$ as in \idents\ and no sum on $j$). The spherical charged
$p$-brane appears and
subsequently expands within the $j$th fluxbrane. To nucleate an uncharged
brane at the intersection of the fluxbranes, one simply adds extra
rotations to $\q'$ as discussed in the previous section.

\newsec{Nucleating Loops of Fundamental String}
\subsec{Test string approximation}

We begin our discussion of fundamental strings by describing the behavior
of a circular test string in flat spacetime coupled to a constant
background $H$ field. Since we are going to consider only classical
solutions, the spacetime can have any dimension larger than two.
We assume that the only non-zero
component of $H$ is $H_{012} = h$ where $h$ is a constant.
The string action is
\eqn\stract{ S= -{1\over 4\pi \a'}
   \int d^2\sigma
\({\sqrt\gamma}\gamma^{ab}\p_\a X^\mu\p_{\beta} X_\mu
+ B_{\mu\nu} \p_\alpha X^\mu \p_\beta X^\nu
\epsilon^{\alpha\beta}\) \ ,}
with $\epsilon^{01}=-1$ and $0\le \sigma\le \pi$.
Choosing the conformal gauge, $\gamma=\eta$, yields the equation of motion
\eqn\eomstr{ \p^2 X_\mu - {1\over 2} H_{\mu\nu\rho} \p_\alpha X^\nu
\p_\beta X^\rho \epsilon^{\alpha\beta} = 0}
and the Virasoro constraints
\eqn\constr{ \dot X^\mu \dot X_\mu + X^{\prime\mu} X_\mu^\prime = 0\ , \qquad
		 \dot X^\mu X_\mu^\prime =0.}
We want to consider solutions describing circular loops, so we set
\eqn\strlop{ X^0 = f(t)\ , \qquad X^1=g(t)\sin2\sigma\ , \qquad
      X^2 = g(t) \cos2\sigma\ , }
with the remaining $X^i$ held constant.

One solution to \eomstr\ and \constr\ is simply $f=2 t/h, \ g=1/h$.
This is a static loop of string with a radius inversely proportional
to the strength of the background $H$ field. It is easy to see that this
solution is unstable: a slightly smaller loop collapses inward, while
a slightly larger loop expands outward. A second solution is
\eqn\accsoln{ f = {2\sin 2t \over h \cos2t} \qquad g = {2\over h \cos2t}\ .}
This describes a loop which initially is twice as large as the static one,
and expands outward. Since $g^2 - f^2 $ is constant, the worldsheet is
a hyperbola, describing constant acceleration. If we analytically continue
in $X^0$ and $t$ we obtain an instanton describing the nucleation of
a loop of string.  The Euclidean action for this instanton is straightforward
to calculate with the result
\eqn\insact{ I = {8\over 3 \a' h^2}}
for all spacetime dimensions. We now construct analogues of these
solutions that include the back-reaction of the string on the spacetime
fields in five dimensions.

\subsec{Spacetime solutions}

Starting with a solution $(g,\phi,F)$
to the equations of motion obtained from \redact\
and performing the duality transformation
\eqn\dual{
\tilde\phi=-\phi,
\qquad F_{D-3}= e^{-4{\sqrt {D-2}\over D-3} \phi}*F,
}
where $F_{D-3}$ is a ($D-3$)-form field strength and the metric $g$ is
left unchanged, we obtain
a ``dual solution" $(g,\tilde\phi, F_{D-3})$
to the equations of motion coming from the action
\eqn\dualact{S={1\over 16\pi G_{D-1}} \int d^{D-1}x \sqrt{- g}\[
R(g) -
{4\over D-3}(\nabla \tilde \phi)^2 -
{2\over (D-3)!} e^{-4{\sqrt{D-2}\over D-3}\tilde  \phi} F^2_{D-3}\]\ .
}
This transformation exchanges magnetic $F$
fields with electric $F_{D-3}$ fields and vice-versa. In the dual variables
there is no longer a connection with Kaluza-Klein theory in $D$ dimensions and
we just have a solution in $(D-1)$ dimensions. Since the metric
is invariant under the duality transformation, all of our previous
solutions can be reinterpreted as the corresponding electric objects.
This is particularly interesting for $D-1=5$ since, as we shall show,
the solution describing the nucleation of a magnetic string in the
last section is transformed into solution describing the nucleation
of a five-dimensional fundamental string.

In the case $D-1=5$, the action \dualact\ is precisely part of
the low-energy effective action of string theory in five dimensions, written
in terms of the
Einstein metric. If we rescale to the string metric
$\tilde g=e^{4\tilde \phi\over 3}g$, this action takes the more familiar form
\eqn\dualac{S={1\over 16\pi G_{5}}\int d^5x \sqrt{-\tilde g}e^{-2\tilde\phi}\[
R(\tilde g) +
4(\nabla \tilde \phi)^2 - {1\over 12}H^2\]\ ,
}
where we have used the notation $H\equiv 2 F_3$. Thus, for every
five-dimensional magnetic solution, there is a dual
electric
solution which extremizes the standard action \dualac. We now discuss some
of these solutions. We will mostly work with the fields appearing in \dualac\
and drop the tildes
on $g$ and $\phi$ for the remainder of this section.

To begin,
recall that the simplest
magnetically charged string in five dimensions was obtained
as the product of a Kaluza-Klein monopole \mon\ with a line.
Transforming to the dual variables, the solution can be re-expressed in
the string frame as
\eqn\fs{
\eqalign{ds^2&=e^{2\phi}(-dt^2+dy^2)+d{\bf x}^2  \ ,\cr
e^{-2\phi}&=1+{4m\over r},\qquad B_{ty}=e^{2\phi}\ ,\cr}}
where $y$ is the coordinate along the line.
 This is the solution corresponding
to the fields about a macroscopic
fundamental string in five dimensions \dghr.

The appropriate background to describe the nucleation of fundamental strings
is given by a uniform electric fluxbrane of dimension two.
This can be constructed
from the $D-1=5$, two-dimensional magnetic fluxbrane discussed in section 2 by
writing it in the dual variables and rescaling to the string metric.
The result is
\eqn\Hftb{
\eqalign{ds^2&=e^{2\phi}(-dt^2+dx_i^2+d\rho^2)+\rho^2d\vp^2\ ,\cr
e^{2\phi}&=1+B^2\rho^2,\qquad H_{t12}=2B\ ,\cr}}
where $(t,x_i)$ $i=1,2$ are coordinates along the fluxbrane. Note that
the  components of the $H$ field are simply constant and that the
induced metric at the center of the
fluxbrane, $\rho=0$, is flat. Since the metric and dilaton both depend
on $B^2$ while $H$ depends linearly on $B$, when $|B|$ is small this solution
reduces to the configuration we started with in the test string discussion
above.

The solution describing the unstable static loop of magnetic string \restsn\
can similarly be dualized. The result is a static loop of fundamental
string in the background $H$ field \Hftb. Since the Einstein metric
is unchanged under duality, the metric \restsn\ also describes a finite
fundamental string loop. This is the exact analog
of the circular test string at rest.

Since the $H$ field has a non-zero
time component, the instanton describing the
nucleation of a loop of fundamental string will have imaginary $H$
as expected for an electric-type field.
It is constructed by starting with the Lorentzian Myers-Perry-Kerr
solution in $D=6$
\eqn\sdkerr{ ds^2 = \(1- {\mu\over r\S}\) d\tau^2 -{2\mu \alpha
 \sin^2\t \over
r\S} d\tau d\vp  + {\S\over r^2 - \a^2-\mu r^{-1}} dr^2 + \S d\t^2}
$$+{\sin^2 \t\over \S} [ (r^2 - \a^2) \S - \mu r^{-1} \a^2 \sin^2 \t] d\vp^2
+r^2 \cos^2\t (d\chi^2 + \cos^2 \chi d\psi^2) \ .$$
We then reduce to five dimensions
using the symmetry $\q'$ of \oneparam\ and \qprime.  Explicitly,
we set $\vp = \tilde \vp + [\Omega -(\sigma /R)] \tau$ and then  read off
the five-dimensional metric, dilaton and gauge field by putting the
metric in the form \red\ with $D=6$ and $x^D = \tau$.  Next we analytically
continue $\psi =- it$. The resulting metric describes an expanding
loop of magnetically charged string. (The Killing vector $\p/\p t$ is
a boost.) We now  apply the duality transformation \dual\ to obtain
an expanding loop of fundamental string. Finally, we analytically
continue back $t=i\psi$ to obtain the desired instanton.

The Euclidean action is not invariant under the duality transformation
\dual. However, for four-dimensional black holes, it has recently been
shown that the rate of pair creating electrically charged black holes
is identical to the rate for creating magnetically charged ones
\refs{\hrdual,\brown}.
This is because one must include a projection onto states of
definite electric charge \preskill\ in calculating the rate which
exactly compensates for the difference in the action. We expect that
a similar result will hold in the present case as well. The rate will
then be given by $e^{-I}$ where $I$ is given by \kerract\ with $p=1$.
In the limit
of small $|B|$, we can use \weakmu\ to express this as
\eqn\finact{ I = {\pi  R \over 6G_5 B^2}\ .}
It was shown in \dghr\ that for a single macroscopic
fundamental string in five dimensions,
the dilaton charge in \fs\ should be given by
$4m=2G_5/\pi\a'$. This can be related to $R$ by recalling that
for the Kaluza-Klein monopole $R=8m$. Using this and setting $h=2B$,
we recover exactly the action found from the test string instanton \insact.

\newsec{Discussion}

We have constructed solutions describing magnetically
charged $p$-branes and loops of fundamental string, as well as
instantons describing the nucleation of these objects in appropriate
background fields. The basic idea was to start with a vacuum solution
with a $U(1)$ isometry. The fixed points of the isometry describe a $p$-brane
in the reduced spacetime which can carry magnetic charge. One can then
apply a duality transformation to obtain electrically charged solutions.
If the reduced
spacetime is five-dimensional, the resulting theory is precisely part of the
low energy string action, and the magnetically charged strings
are transformed into fundamental strings.

To construct our solutions we have always started with a Euclidean
black hole or Euclidean black hole cross time.
However, it is clear that there are many other
possibilities which can yield interesting solutions.
For example, one can start with a Lorentzian black hole
cross a circle\foot{If the radius of the circle is
small compared with the mass of the black hole then this solution
is likely to be stable \refs{\gl}.}.
If one considers the symmetry consisting of translation
around the circle plus rotation of the black hole, the reduced space
describes a black hole in a background magnetic fluxbrane. This is the
likely endpoint of a $p$-brane which is smaller than the static radius
and collapses to form a black hole. In five dimensions, we can dualize
to obtain a black hole in a background $H$ field.

It was shown in \decay\ that in the standard five-dimensional
\KK\ theory, the dominant decay mode for the  weak magnetic fields
of physical relevance
was via  ``bubble nucleation" analogous to the decay of the \KK\
vacuum.
The same is true for the decay of the
magnetic fluxbranes described here. Indeed the earlier analysis
is just the special case $p=0$. For every $p$, the reduction
of the $(p+5)$-dimensional Myers-Perry-Kerr  instanton,
with one non-zero rotation parameter, via $\q$ \oneparam\
describes decay of a $(p+1)$-fluxbrane via ``bubble nucleation"
while the shifted reduction via $\q'$ \qprime\
describes nucleation of a charged $p$-brane.
If we take two instantons, with different parameters,
one reduced along $\q$ and the other along $\q'$,
so that the asymptotic
value of the magnetic field $B$ is the same in both cases, then we find
that the bubble nucleation has smaller action for small $|B|$.

Having  said this, it was also pointed out in \decay\ that a
spin structure argument analogous to that
which would stabilize the $D=5$ \KK\ vacuum \witten\
would also rule out the bubble nucleation but allow the pair production
of monopoles. Roughly the argument is that the
$D=5$ \KK\ Melvin solution admits two spin structures which may be
distinguished by asking what phase
spinors pick up under parallel transport around
the internal circle. There are two instantons for the decay of the same
four-dimensional flux tube; one corresponding to bubble nucleation and the
other to pair production. Each instanton
admits only a single spin structure since it
is simply connected, but that spin structure tends at infinity
to a different one of the two possibilities. So depending on which
spin structure is chosen for the background Melvin spacetime, one or other
of the decay channels is ruled out. If $|B|$ is small then
there is a natural
choice which is in some sense continuous with the choice that rules out
the vacuum decay. This allows pair production but eliminates the
bubble nucleation. In particular, this is what we expect in
$S^1$ compactifications that preserve supersymmetry.

A similar argument can be used to show that there is again a natural choice of
spin structure for a single background $(p+1)$-fluxbrane which
would rule out the
decay via bubble nucleation but allow the decay via production of
a spherical charged $p$-brane. These are the only
two possible decay routes. When the background is a configuration
of intersecting fluxbranes there are more possibilities for
the decay. Suppose we have $k$ intersecting fluxbranes in
$p+4$ spacetime dimensions. One
decay channel that always exists is the bubble nucleation. Then there
are $k$ channels which are the nucleation and subsequent
expansion of a charged $p$-brane within  each individual fluxbrane.
And $\(\matrix{k\cr 2}\)$ channels which correspond to an uncharged
$(p-2)$-brane produced and
expanding in the intersection of each pair of fluxbranes and so on:
 $\(\matrix{k\cr l}\)$ possible  $(p-2l+2)$-branes produced in the
intersection of each subset of $l$ fluxbranes.
The generalization of the spin structure argument seems
to result in the bubble nucleation being ruled out, the
$(p-4n)$-brane production being allowed, where $n$ is an integer, and
the $(p-2(2n+1))$-brane production being ruled out.  So, for example,
Bais-Batenberg ``monopole'' pair production would be allowed only
if the reduced spacetime dimension were a multiple of 4.

The situation with the fundamental string is slightly different. After
we dualize in five dimensions, the connection with six dimensions is lost
and in particular we no longer have a spin structure argument.
It seems that there should be an argument to eliminate the
dual of the bubble nucleation process, whilst keeping the string production
process, and the following is a promising possibility.
While the instanton describing bubble nucleation is non-singular in
six dimensions, it is singular in five, as is its dual.
It is not clear whether this singular dual instanton
corresponds to a physical decay channel of the $H$ field but we expect not
since if it is allowed it suggests that the vacuum itself would also
decay via dual-bubble nucleation.
Of course, the
instanton describing the nucleation of  a fundamental string is also singular,
but here the singularity is readily interpreted in terms of the string
source and almost certainly should be allowed.

The extended objects that we have considered are all extremal, in the
sense that their mass per unit $p$-volume was essentially equal to their
charge. It would be interesting to know whether one could nucleate
nonextremal extended objects. For  black holes in four
dimensions, it was found that nonextreme black holes were created in thermal
equilibrium with their Hawking temperature equal to the acceleration
temperature. It thus seems that a necessary condition to nucleate a nonextremal
$p$-brane is that the
Hawking temperature  must go to zero in the extremal limit, so that it
can equal the acceleration temperature for small acceleration.

Our construction has only yielded fundamental strings in five dimensions.
However, from the test string calculation it is clear that there should
exist analogous solutions in all dimensions larger than two. In particular,
a four-dimensional solution describing a loop of fundamental string
should exist. It would
be interesting to find it.

In addition to the fundamental strings,
we have mentioned that our solutions have two other string theory
interpretations. Firstly, the \KK\ reduction of D=11 leads to 6-brane solutions
carrying Ramond-Ramond charge of the D=10 type IIA theory.
On the other hand, for $D\le 10$, the \KK\ solutions provide solutions
to string theory compactifications which include an $S^1$ factor. Since these
are charged with respect to the $U(1)$ gauge field coming from the metric,
they carry NS-NS charge in the type II theory.
Let us briefly mention some ways in which we can generalise our solutions.
For convenience we discuss these  transformations in terms of the
simplest flat charged $p$-branes.
By wrapping the type IIA 6-brane solution around an $n$-torus we can obtain
$(6-n)$-brane solutions of type II theory
in $(10-n)$ dimensions that carry RR charge. These will be
related to the NS-NS $(6-n)$-branes obtained by Kaluza-Klein reduction
by some field redefinitions (part of the continuous group of $U$-duality
transformations \hulltown).
Another way to obtain new solutions is to
use the fact that the \KK\ solutions have a $U(1)$ isometry.
In particular, there is the T-duality
symmetry which includes interchanging the two $U(1)$'s coming from the
dimensional reduction of the metric and antisymmetric tensor.
This transformation takes
the ``metric" $p$-brane in $p+4$ reduced spacetime dimensions
to an ``anti-symmetric
tensor" $p$-brane. These latter objects can be considered to be $H$-monopoles
in four dimensions \refs{\ghl,\khuri}
with $p$ flat dimensions added. Equivalently, the
$p=5$ solution in nine dimensions can be constructed by taking
a periodic array of 5-branes in ten dimensions to get a 5-brane in nine
dimensions. By wrapping these solutions around an $S^1$
we can then obtain a 4-brane in eight dimensions etc. Finally, new solutions
can also be obtained by employing various string-string dualities,
which amounts
to writing the solutions in suitable dual variables \refs{\hulltown,\wittena}.
It is natural to expect that all of the
above
transformations acting on our instantons will produce instantons
describing the nucleation
of the corresponding objects.

In recent work Polchinski has shown that
$D$-branes, surfaces where first quantised
strings have Dirichlet boundary conditions,
are carriers of Ramond-Ramond charges \polchinski. In particular, the 6-brane
of the type IIA theory has a $D$-brane description.
This identification has, as yet,
only be made in the static, supersymmetric case. Although our instantons
are neither static nor supersymmetric, we still might expect a
related $D$-brane construction. Having such a construction might enable
one to go beyond the semi-classical approximation in a controlled manner.

\bigskip

\centerline{\bf Acknowledgements}
We would like to thank David Kastor for useful discussions
and the Aspen Center for Physics for
hospitality during the initial stages of this work.
FD and JPG are supported by the U.~S.~Department of Energy
under Grant No. DE-FG03-92-ER40701.
GTH was supported in part by NSF Grant PHY95-07065.

\appendix A ~

We here calculate the action of the Euclidean rotating black hole
with one non-zero rotation parameter in arbitrary dimension.
The metric is \mp
\eqn\eucmp{
 ds^2 = \(1- {\mu\over r^{D-5}\S}\) d\tau^2 -{2\mu \alpha
 \sin^2\t \over
r^{D-5}\S} d\tau d\vp  + {\S\over r^2 - \a^2-\mu r^{5-D}} dr^2 + \S d\t^2}
$$+{\sin^2 \t\over \S} [ (r^2 - \a^2) \S - \mu r^{5-D} \a^2 \sin^2 \t] d\vp^2
+r^2 \cos^2\t d\Omega_{D-4} \ ,$$
where $\S = r^2 - \a^2\cos^2\t$. The horizon is located at $r = \r$ where
\eqn\apphor{ \r^2 = \a^2 + {\mu\over \r^{D-5}} \ .}
The radius of the circle at infinity is
\eqn\apprad{
R = {1\over \kappa} = {2\mu\r^{6-D} \over (D-3)\r^2 - (D-5) \a^2}\ ,}
while the Euclidean angular velocity is
\eqn\appang{ \Omega = {\a \r^{D-5}\over \mu}\ .}

The Euclidean action is defined with respect to a background geometry:
\eqn\appact{I = - {1\over 16\pi G_D} \int d^Dx \sqrt{-g_D}{}\ R(g_D)
                - {1\over 8\pi G_D}\int d^{D-1}x \sqrt{h}(K-K_0)\ ,}
where $K$ is the trace of the extrinsic curvature of the boundary and
$K_0$ is the trace of the extrinsic curvature of the boundary embedded
in the background geometry. Here the appropriate background is just
flat ${\R}^D$.

The instanton is Ricci flat so the boundary term is the only contribution.
Let the boundary be given by $r=$ constant. The induced metric is
\eqn\induced{\eqalign{
ds_{D-1}^2 =  & \(1- {\mu\over r^{D-5}\S}\) d\tau^2 -{2\mu \alpha
 \sin^2\t \over r^{D-5}\S} d\tau d\vp + \Sigma d\t^2 \cr
& +  {\sin^2 \t\over \S} [ (r^2 - \a^2) \S - \mu r^{5-D} \a^2 \sin^2 \t] d\vp^2
+r^2 \cos^2\t d\Omega_{D-4}\ . \cr}
}
The determinant is
\eqn\appdet{\eqalign{
\sqrt{h} =&\(1 - {\a^2\over r^2} - {\mu\over r^{D-3}}\)^{\half}
\(1-{\a^2\over r^2}\cos^2\t\)^{\half} \cr
&r^{D-2}\sin\t\cos^{D-4}\t \sqrt{\Omega_{D-4}}\ .\cr
}}
The unit normal is
\eqn\appnorm{n = \({r^2-\a^2 -\mu r^{5-D} \over r^2 - \a^2 \cos^2\t}\)^{\half}
                {\p\over\p r}\ .
}
$K$ is calculated via
$K\sqrt{h} = n\sqrt{h}$. So that
\eqn\appk{K = {n \sqrt{h} \over \sqrt{h}}\ .
}
The background value $K_0$ is easily computed from this by setting
$\mu=0$ since \eucmp\ with $\mu$ zero is flat for all values of
$\alpha$. Thus, we have
\eqn\integrand{
\(K - K_0\) \sqrt{h} = n\sqrt{h} - n \sqrt{h}\vert_{\mu=0}
{\sqrt{h}\over \sqrt{h}\vert_{\mu=0}}\ .
}
We want to take the limit $r\rightarrow \infty$. In this limit
\eqn\applim{
\lim_{r\to \infty}{\({\sqrt{h}\over \sqrt{h}\vert_{\mu=0}}\) }=
1-{\mu\over 2r^{D-3}}\ ,
}
and hence,
\eqn\morelim{\eqalign{
\lim_{r\to \infty}\[ (K-K_0) \sqrt{h}\] = &
\lim_{r\to\infty}\[ (n\sqrt{h} - n\sqrt{h}\vert_{\mu=0})
+ n\sqrt{h}\vert_{\mu=0}{\mu\over 2 r^{D-3}} \]\cr
 = & \lim_{r\to \infty}\[ {\p n\sqrt{h} \over \p \mu}\Big\vert_{\mu=0}
+{1\over 2 r^{D-3}} n\sqrt{h}\vert_{\mu=0} \]\mu \cr
= & - \half \mu \sin\t \cos^{D-4}\t \sqrt{\Omega_{D-4}}\ .\cr}
}
Then
\eqn\calc{
I = {\pi RV_{D-4}\over 4 (D-3)G_D} \mu =
{V_{D-4}\over 8(D-3)G_{D-1}} \mu\ ,
}
where $V_{D-4}$ is the volume of the unit $(D-4)$-sphere.

We can check that this agrees with the thermodynamic and Smarr
formulae given in \refs{\mp}. The mass, $M$, and angular momentum, $J$,
are given by
\eqn\massang{
M= {(D-2) V_{D-2} \over 16 \pi G_D} \mu, \qquad
J = {2\over D-2} Ma\ ,
}
where $a=i\alpha$ is the Lorentzian rotation parameter.
The Smarr relation is
\eqn\smarr{ \omega J + TS = {D-3\over D-2} M \ ,}
where $S$ is the entropy, $T = \kappa/2\pi$ the temperature,
and $\omega$ is the Lorentzian angular velocity.
The thermodynamic potential $W$ is
\eqn\thermpot{ W = M -TS -\omega J = {1\over D-2} M \ ,}
and thus the Euclidean action is
\eqn\wovert{I = {W\over T} = {2\pi R  \over D-2} M \ ,}
which agrees with  \calc\ since $V_{D-2} = 2 \pi V_{D-4}/(D-3)$.

It is more interesting to express \calc\ in terms of the
value of the asymptotic magnetic field strength, $B =  \Omega-
(\sigma/R)$ ($\sigma=\Omega/|\Omega|$), and
radius of compactification, $R$.  From \apphor\ and \appang, we can
eliminate $\mu$, and then solve for $\a$ with the result
\eqn\alphars{ \a = {-1 +\sqrt{ 1+ 4r^2_H \Omega^2} \over 2\Omega} \ .}
Using this and \appang\ in \apprad\ we obtain an expression for $R$ in
terms of $\r$ and $\Omega$. This can be inverted to yield
\eqn\raitch{
{r_H\over R} = {
 D -4 + \[(D-4)^2 - (1-R^2\Omega^2)
(D-3)(D-5) \]^{1/2} \over  2(1-R^2 \Omega^2)}\ .}
 One can thus obtain an expression for $\mu = \alpha r_H^{D-5}/\Omega$ in
 terms of $R$ and $\Omega$ which unfortunately is extremely complicated.
 However, it simplifies in two limits.
When $\alpha =0$, the instanton is just the  Euclidean Schwarzschild
solution and  one has
\eqn\schwact{
\mu =\[{(D-3) R\over 2}\]^{D-3}\ .
}
When $|\Omega R| \approx 1$ (so $|B|$ is small) one finds
\eqn\smalb{ \mu = \( {D-4\over 2}\)^{D-4} {R\over |B|^{D-4}} \ .}

\listrefs
\end